\documentclass[11pt]{article}

\usepackage{authblk}  % have affiliations with article
\usepackage{graphicx} % Include figure files
\usepackage{epsfig}
\usepackage{dcolumn}  % Align table columns on decimal point
\usepackage{bm}       % bold math
\usepackage{amsmath}
\usepackage{amssymb}
\usepackage{url}

\usepackage{float}

\usepackage{multirow}
\usepackage{booktabs}

\newcommand{\beq}{\begin{equation}} \newcommand{\eeq}{\end{equation}}
\newcommand{\bea}{\begin{eqnarray}} \newcommand{\eea}{\end{eqnarray}}

  \newcommand
{\Romannumeral}[1]{\uppercase\expandafter{\romannumeral#1}}

\newcommand{\be}{\begin{enumerate}} \newcommand{\ee}{\end{enumerate}}
\newcommand{\bi}{\begin{itemize}} \newcommand{\ei}{\end{itemize}}
\newcommand{\ba}{\begin{array}} \newcommand{\ea}{\end{array}}
\newcommand{\bc}{\begin{center}} \newcommand{\ec}{\end{center}}
\newcommand{\bt}{\begin{tabular}} \newcommand{\et}{\end{tabular}}

\def\lsim{\mathrel{\rlap{\lower4pt\hbox{\hskip1pt$\sim$}}
    \raise1pt\hbox{$<$}}}           % less than or approx. symbol
\def\gsim{\mathrel{\rlap{\lower4pt\hbox{\hskip1pt$\sim$}}
    \raise1pt\hbox{$>$}}}           % greater than or approx. symbol

\newcommand{\half}{\textstyle {1\over2} \displaystyle}    % One half
\newcommand{\Dslash}{{\hbox{D}\kern-0.6em\raise0.15ex\hbox{/}}} % D slash

\renewcommand{\et}{\eta}

\begin{document}
	\thispagestyle{empty} % suppresses display 1st page's number
	
	\setlength{\oddsidemargin}{0cm}
	\setlength{\baselineskip}{7mm}

\begin{normalsize}\begin{flushright}

% CERN-PH-TH/2006-147 \\
% DAMTP-2006-59 \\
% UCI-2004-xx \\
July 2018 \\

\end{flushright}\end{normalsize}

\begin{center}
  
\vspace{15pt}

{\Large \bf Gravitational Fluctuations as an Alternative to Inflation}

\vspace{30pt}

{\sl Herbert W. Hamber\footnote{HHamber@uci.edu.}, Lu Heng Sunny Yu \footnote{Lhyu1@uci.edu.}} 
\\
Department of Physics and Astronomy \\
University of California \\
Irvine, California 92697-4575, USA \\

\vspace{10pt}
\end{center}

%  ABSTRACT 

% XXXXXXXXXXXXXXXXXXXXXXXXXXXXXXXXXXXXXXXXXXXXXXXXXXXXXXXXXXXXXXXXXXXXXXXXXX

\begin{center} 
	{\bf ABSTRACT } 
\end{center}

\noindent

The ability to reproduce the observed matter power spectrum $P(k)$ to high accuracy is often considered as a triumph of inflation.
In this work, we explore an alternative explanation for the power spectrum based on nonperturbative quantum field-theoretical methods applied to Einstein's gravity, instead of ones based on inflation models. 
In particular the power spectral index, which governs the slope on the $P(k)$ graph, can be related to nontrivial scaling exponents derived from the Wilson renormalization group analysis.  
We find that the derived value fits favorably with the Sloan Digital Sky Survey telescope data. 
We then make use of the transfer functions, based only on the Boltzmann equations which describe states out of equilibrium, and Einstein's General Relativity, to extrapolate the power spectrum to the Cosmic Microwave Background (CMB) regime.  
We observe that the results fit rather well with current data. 
Our approach contrasts with the conventional explanation which uses inflation to generate the scale invariant Harrison-Zel'dovich spectrum on CMB scales, and uses the transfer function to extrapolate it to galaxy regime. 
The results we present here only assume quantum field theory and Einstein's Gravity, and hence provide a competing explanation of the power spectrum, without relying on the assumptions usually associated with inflationary models.
At the end, we also outline several testable predictions in this picture that deviate from the conventional picture of inflation, and which hopefully will become verifiable in the near future with increasingly accurate measurements.

%%% Published in : Universe 2019, 5(1), 31; https://doi.org/10.3390/universe5010031 
%%% DOI  : 10.3390/universe5010031

%%% Keywords : Quantum Cosmology, Quantum Gravity, Inflationary Cosmology

\newpage

% XXXXXXXXXXXXXXXXXXXXXXXXXXXXXXXXXXXXXXXXXXXXXXXXXXXXXXXXXXXXXXXXXXXXXXXXXX

\section{Introduction}
\label{sec:intro}  

\vskip 10pt

In cosmology, we know that matter in the universe is not homogeneously and isotropically distributed.  Nevertheless, these inhomogeneities are far from random, but congregated in a rather specific manner.
Detailed measurements of these fluctuations have been made through galaxy and cosmic microwave background (CMB) surveys, and find that they can be characterized by an almost scale invariant correlation function referred to as the power spectrum.  
The questions of how and why matter is distributed the way it is are thus important ones in cosmology.  

One conventional explanation is the theory of cosmic inflation involving a scalar field called the inflaton.  Inflation explains that matter fluctuations are originated from quantum fluctuations of this scalar field, and is able to reproduce the observed power spectrum of matter fluctuation to a high degree of accuracy.  This is known as one of the earliest quantitative verification of inflation, and often considered as a triumph of inflation.
Nevertheless, the detail model of inflation is still largely unknown, and many models suffer from fine tuning problems.
As a result, the goal of our paper is to explain the shape of the power spectrum independent of inflation.  
In this work we point out that using well-established nonperturbative quantum field theory techniques, one can deduce the scaling of the correlation functions of macroscopic gravitational fluctuations, and these fluctuations precisely produces the shape of the matter power spectrum.
It should be pointed out that an explanation of the matter power spectrum independent of inflation as presented here is, to our knowledge, a first of its kind.

Given the vastness of the subject, we should clarify that it is impossible address all the cosmological problems raised by inflation at once, nor is this the intent of our paper.  
Hence the paper will focus on the central problem of explaining the matter power spectrum.
As a result, the scope of our paper is rather restricted and two-fold.  
(i) To describe this perspective and present detailed calculations of the power spectrum as reproduced from macroscopic gravitational fluctuations, and (ii) to describe predictions of deviations from the classical picture, which serves as a test for this gravitational picture as experimental accuracies in the large-scale regions improve in the near future.  
Nevertheless, the other cosmological problems addressed by inflation, including an explanation of why the universe is flat, isotropic and homogeneous, remains of great interest.  
This, amongst other inflation related problems, such as B-mode polarizations and quantitative tests against higher order correlation functions, are intended to be pursued in future work.  

According to current established physical laws, two main aspects determine the cosmological evolution. The first aspect involves the use of the correct field equations for a coupled matter gravity system, and these in turn follow from Einstein's classical field equations, coupled with an exhaustive specification of all the various matter and radiation components and their mutual interactions. 
The second main ingredient is a set of suitable initial conditions for all the fields in question, which should then lead to a hopefully unambiguous determination of the complete subsequent time evolution.

While there is currently very little disagreement on the nature of the correct cosmological evolution equations themselves, which largely follow from the choice of a suitable metric based on physical symmetry arguments and on a complete list of matter and radiation constituents based on our current understanding of fundamental particle physics, the same cannot be said for the choice of initial conditions.
The latter are largely unknown, and generally involve a number of explicitly stated, and sometimes implicitly assumed, assumptions about what the universe might have looked like close to the initial singularity.  Furthermore, it is generally expected that quantum effects will play a major role at such early stages, be it for the matter fields and their interactions, or for the gravitational field itself for which a classical description is clearly inadequate in this regime. 
Indeed, over the years attempts have been made to partially include some quantum effects, by assuming non-trivial Gaussian (free field) correlations for some matter and gravity two point functions.

Recent attempts at overcoming our admitted ignorance regarding the initial conditions for the evolution equations have followed a number of different avenues. 
In one popular scenario \cite{ven02,ven04,ven07}, it is argued that the universe might have evolved initially from a state that could not even remotely be described in terms of a simple and symmetric physical characterization.  
If, as suggested by particle physics and string theory, a modified quantum dynamics becomes operative at short distance, then one would expect a complete removal of the initial spacetime singularity, replaced here instead by some sort of bounce. 
One key, but nevertheless natural, consequence of this perspective is that the universe must have evolved into and out of the initial singularity in a highly coherent quantum state, with non-trivial quantum correlations arising between all fields, with the latter presumably operating on all distance scales. 
These highly coherent and complex initial conditions would then represent the surviving physical imprint of a previous, presumably very long, cycle of cosmological evolution, thus generating a set of suitable initial conditions for the cosmology of our current universe.

Another possibility regarding the initial conditions is inspired by the semi-classical approach to quantum gravity \cite{har83,har08} , and follows from a more geometric point of view regarding the nature of gravity. 
In this approach the universe naturally evolves instead from an initial simple, symmetric and elegant geometric construct, as described in practice for example by the so-called no-boundary proposal for homogeneous isotropic closed universes, endowed with a cosmological constant and a variety of other fields.
Another popular attempt to explain features of the early universe, and more specifically properties of the matter power spectrum, is through inflation models \cite{gut81,lin82,alb82}.
These propose that structures visible in the Universe today get formed from quantum fluctuations in a hypothetical primordial scalar inflaton field.  
Although the precise behavior and dynamics of the inflaton field are to this day still largely controversial \cite{ste11,ste12,teg05,ste13,ste14,ste17}, one nevertheless finds some general features that are common among these models, and which allow one to derive predictions about the nature of the power spectrum.  

In broad terms, the approach followed in this paper can be viewed as more in line with the first scenario, where as little as possible is assumed about the nature of the initial quantum state of the universe, given that, as mentioned previously, the latter might be quite far from a simple and 
symmetric physical characterization. 
Instead, in this paper, we make extensive use of previously gained knowledge from nonperturbative studies of quantum gravity regarding the large distance behavior of gravitational and matter two-point functions.  
An important ingredient in the results presented below will be therefore the non-trivial scaling dimensions obtained from these studies, and how these relate to current observational data.
We note here that the quantum field theoretic treatment of perturbatively non-renormalizable
theories, and the determination of their generally non-trivial scaling dimensions, has a rather distinguished history, originally developed in the context of scalar field theories by Wilson and Parisi \cite{wil72,par73,par76}, and recently reviewed in great detail for example in \cite{zin02}, 
as well as in many other excellent monographs \cite{par81,itz91,car96,bre10}.
Attempts at quantizing gravity within established rules of quantum field theory also have a long and distinguished history, dating back to Feynman's original investigations, and the subsequent 
formulation of a consistent covariant path integral framework \cite{fey63,fey95}.
Quantum gravity, and by now the rather compelling theoretical evidence of a nontrivial ultraviolet (UV) renormalization group fixed point in four spacetime dimensions, in principle leads to a number of unambiguous predictions, recently reviewed and summarized in \cite{book,ham17}.
Perhaps the most salient observational effects of such an approach include a running of Newton's constant $G$ with scale on very large cosmological distances \cite{hw05}, the modification of classical results for the growth of relativistic matter density perturbations and their associated growth exponents, and a non-vanishing slip function in the conformal Newtonian gauge \cite{rei10}.
Moreover, the existence of a non-trivial UV fixed point, arising from the highly non-linear nature of the Einstein-Hilbert action, leads in a natural way to a nontrivial quantum condensate in the curvature.  
Such a condensate effectively produces long-distance correlations between local curvature fluctuations and, through the field equations, these couple locally to matter density fluctuations.  
The fluctuations in matter density, which are in principle unambiguously calculable in this perspective, should lead to gravitational clumping of matter in the over-dense areas, which eventually brings about
the formation of galaxies.  
In other words, the correlations of galaxy distributions should in principle become predictable from correlations of curvature fluctuations derived from quantum gravity.  
The calculated results can then be compared with observations, and thus viewed as a potential test for the proposed vacuum condensate picture of quantum gravity.
\footnote{
A possible quantum-mechanical origin of cosmological fluctuations in the currently observed galaxy matter distribution, as well as in the cosmic microwave background, was first proposed in \cite{muk81}.
}

The paper is organized as follows.  
In Section 2 we outline the canonical ways to parameterize matter distribution quantitatively in cosmology, via $n$-point correlation functions and the power spectrum, which in turn are largely characterized by the so-called spectral indices.  
After introducing and defining the correlation functions, we end the section by quoting observed values of the spectral indices from modern measurements.  
Subsequently, in Section 3 we provide an explanation for the origin of matter fluctuations as a result of underlying quantum gravitational fluctuations for which the correlation functions, and hence spectral indices, can then be unambiguously calculated.  
We will then show that the theoretical predictions are in good agreement with the observed results.  
Section 4 discusses how this theoretical prediction can be extended via the so-called transfer function to the earlier time of a radiation-dominated epoch, which again shows a generically good agreement with current CMB and other measurements.  
Section 5 outlines the alternative, more conventional, explanation of the same spectrum via inflation, and briefly discusses the differences between the classical and the quantum gravity picture.  
Section 6 elaborates on further physical effects predicted by the quantum theory of gravity.  
Specifically, the vacuum condensate picture of quantum gravity suggests both the existence of an effective infrared (IR) cutoff (related to the observed cosmological constant) and a weak renormalization group (RG) running of Newton's $G$ on very large scales.  
Here we present how these corrections will affect the power spectrum, and discuss how these effects can be used to distinguish the quantum gravity picture from the classical inflationary picture, with increasingly precise data becoming available in the near future. 
Section 7 summarizes the key points of our study and concludes the paper.

\vskip 20pt

\section{Background -- Correlation Functions and Power Spectrum}
\label{sec:introPk} 

\vskip 10pt

In this section, and as a background, we briefly summarize the two main quantities that are used in cosmology to quantitatively parameterize matter fluctuations: (i) correlation functions in both real and momentum space, and (ii) their corresponding powers, or spectral indices.
To begin, it is customary to describe the matter density fluctuations in terms of the matter density contrast $\delta({\bf x},t)$ \cite{pee93,wei08}, which measures the fractional over-density of matter relative to the background mean density $\bar{\rho}(t)$,
\begin{equation}
\delta({\bf x},t) \equiv \frac{\rho({\bf x},t)-\bar{\rho}(t)}{\bar{\rho}(t)} = \frac{\delta\rho({\bf x},t)}{\bar\rho(t)},
\label{eq:delta_def} 
\end{equation}
where $\bar\rho(t)$ evolves only with time, in a way that is governed by the background Friedman-Robertson-Walker (FRW) metric. 
The most common measure of such fluctuations is the real space correlation function $G_\rho(r)$, which averages the density contrast over all space 
\footnote{The galaxy matter density correlation function in real space is usually denoted as $\xi_M$, or simply $\xi$, in the cosmology literature.  
But in this paper it will be desirable to call it $G_\rho$, to avoid confusion with the quantum gravitational correlation length $\xi$ that will be introduced later.}
\begin{equation}
G_\rho({\bf x},t;{\bf x'},t') 
\equiv 
\langle \delta({\bf x},t) \; \delta({\bf x'},t') \rangle 
= \frac{1}{V} 
\int_V   d^3{\bf y}
 \; \delta({\bf x+y},t) \; \delta({\bf x'+y},t') \;\; 
\label{eq:G_rho_def} 
\end{equation}
or its momentum space conjugate, commonly known as the power spectrum
\begin{equation}
P(k) = (2\pi)^3 \langle \; |\delta(k,t_0)|^2 \; \rangle 
= (2\pi)^3 F(t_0)^2 \langle \; | \Delta(k)| ^2 \; \rangle \;\; .
\label{eq:Pk_def}
\end{equation}
In the second expression the time dependence is factored out as
\footnote{
$F(t)$ simply evolves according to the FRW background;  its explicit evolution can be found in \cite{wei08}.} 
$\delta(k,t)\equiv\Delta(k)F(t)$, so that the power spectrum explicitly compares fluctuations as they are measured today, at time $t=t_0$.  Furthermore, both the real and momentum space correlation functions represent statistical averages, which satisfy homogeneity and isotropy, and hence are expected to only depend on the magnitude $r\equiv|\mathbf{x-y}|$ and $k$.  

The power spectrum is directly related to the real space correlation function via the Fourier transform
\beq
G_\rho(r;t,t') = \frac{1}{2\pi^2} \frac{F(t)F(t')}{F^2(t_0)} 
\cdot 
\int_{0}^{\infty} dk \; k^2 \; P(k) \; \frac{\sin(kr)}{kr} \;\; .
\label{eq:G_rho_FT_Pk}
\eeq
In particular, most galactic observations are taken at relatively low redshifts, where 
$t\approx t'\approx t_0$, in which case the pre-factor ratio of $F(t)$'s reduces to unity.
One often starts with so-called scale-invariant models in cosmology, which assume a simple power law for the power spectrum, characterized by scaling indices.  
The real-space correlation function is then usually parameterized by a single index $\gamma$ \cite{pee93}, defined via
\beq
G_\rho(r)= \left( \frac{r_0}{r} \right)^\gamma \;\; .
\label{eq:gamma_def}
\eeq
In momentum space, the spectral index $s$ is then defined via
\beq
P(k)=\frac{a_0}{k^s}  \;\; .
\label{eq:s_def}
\eeq
With this parametrization, the Fourier transform can be evaluated through (\ref{eq:G_rho_FT_Pk}), or formally and exactly in $d$ dimensions
\begin{equation}
	\int d^d x \; e^{-i k \cdot x} \frac{1}{x^{2n}}=
	\frac{\pi^{d/2} 2^{d-2n} \Gamma(\frac{d-2n}{2})}{\Gamma (n)} \frac{1}{k^{d-2n}}
	\label{eq:FTexact} \; \; ,
\end{equation}
which then gives the relation
\beq
\gamma=3-s \;\; .
\label{eq:gamma_3-s} 
\eeq 
These scale-invariant models have been compared to increasingly accurate astrophysical data measurements over the past few decades.  
Earlier data \cite{pee93} supported this model with approximate values $\gamma=1.77 \pm 0.04$ 
and $ r_0 = 5.4 \pm 1 \; h^{-1}\text{Mpc} $ for distances in the range 0.01 
$h^{-1}$Mpc $\lesssim r \lesssim$ 10 $h^{-1}$Mpc, which then gives for the exponent $s \simeq 1.23$.  
More recent data \cite{bah03,pee98} again supports this model, with approximate values
$ \gamma = 1.8 \pm 0.3 $ for distances in the range 
$ 0.1 h^{-1}$Mpc $\lesssim r \lesssim$ 50 $h^{-1} $Mpc, 
which leads to $s \simeq 1.2$.  
Another set of recent estimates gives $\gamma = (1.79,1.84)$  for distances up to 100Mpc, 
see for example \cite{bau06,lon07,teg02,teg04,dur14,wan13,coi12} and references therein.
Here we will focus on the most recent high statistics data from Sloan Digital Sky Survey (SDSS) collaboration \cite{gil18},
which supports the above model with approximate values
$ s = 1.089 $ (and thus $ \gamma \simeq 1.91 $) for distances in the range 
$ 0.01 \; h \; \text{Mpc}^{-1} \lesssim k \lesssim 0.3 \; h \; \text{Mpc}^{-1} $.  
The latest SDSS observational measurements will later be shown here in the next section.
\footnote{
For a comprehensive list of commonly accepted cosmological parameters and their uncertainties, we refer the reader to the recent authoritative reference \cite{dam06}.}

An important challenge in cosmology is to provide a theoretical explanation for these particular values
of $s$ and $\gamma$.  
Here it should be noted that correlation functions for matter fields can be related to that of gravitational fields via the Einstein field equations, and the latter are in principle calculable from a quantum field theoretical treatment of gravity.  
Therefore quantum gravity provides a natural framework for calculating and explaining the scaling exponent $\gamma$ as well as the spectral index $s$.  
This is the general picture we will discuss in detail in the following section.

\vskip 20pt

\section{Quantum Gravity effects as an explanation of $P(k)$}
\label{sec:QGexpofPk}  

\vskip 10pt

The previous section outlined the customary ways to quantify matter fluctuations via the correlation functions $G_\rho (r) $ and $P(k)$, and the corresponding spectral indices $\gamma$ and $s$.  
Here we present a possible explanation for the origin for matter fluctuations as a direct result from gravitational fluctuations, and show how the corresponding correlation functions can be calculated from first principles from a quantum-field treatment of gravity.  
In the end we will show how the theoretically predicted results are in surprisingly good agreement with the most recent galaxy data from the SDSS telescope collaboration \cite{gil18}. 

General relativity provides an unambiguous relation between curvature and matter distributions through the Einstein field equations
\footnote{
Here we ignore the cosmological constant $\lambda$ term for now, since we do not expect dark energy to be significant in the epoch of matter and galaxy formation, nor should it have significant effects on fluctuations due to its homogeneous nature.}
\beq
R_{\mu\nu} \, - \, \half \, g_{\mu\nu} R \, = \, 8\pi \, G \; T_{\mu\nu} \;\; ,
\label{eq:EFE}
\eeq
and it is therefore natural that any fluctuations in matter are directly related to fluctuations in curvature.  
Now, in any quantum theory, quantities fluctuate, and a quantum theory of gravity produces metric and curvature fluctuations for which correlation functions are in principle calculable from first principles.  
It should then be possible to relate unambiguously such correlation functions to matter fluctuations, via the Einstein field equations.  
Here we will first outline briefly how invariant curvature correlation functions can be calculated from a quantum theory of gravity \cite{cor94}.  
Secondly, we will show how matter correlation functions can be determined from curvature correlations, and at the end we will compare these predictions to observational data. 

Quantum gravity, like QED and QCD, is in principle a unique theory.  
In the Feynman path integral approach, only two key ingredients are needed, namely, a gravitational action and a functional measure over metrics.  
As shown originally by Feynman \cite{fey63, fey95}, the Einstein-Hilbert action plus a cosmological constant term represents the unique action to describe a massless spin-2 field.  
Additional higher derivative terms are in principle consistent with general covariance, but only affect the physics at very short distances, and therefore will not be considered here, since we are mainly interested in large distance cosmological physics.  
From this perspective, the formulation of a quantum theory of gravity in this approach has essentially no adjustable parameters, in a way that is quite similar to QCD and non-Abelian gauge theories in general.

A special characteristic of quantum gravity is its perturbative non-renormalizablility and its highly non-linear nature.  
As in the case of other non-linear theories such as QCD, Yang-Mills theories, and the $O(N)$ non-linear sigma model, one can nevertheless extract universal quantities, such as critical exponents (such as the exponents $\nu$, $\gamma$, $\delta$ etc. in the nonlinear sigma model), and genuinely nonperturbative characteristic scales (such as correlation lengths or $\Lambda_{\overline{MS}}$ for QCD).  
Such predictions, though obtained from a nonpertubative treatment, are today amongst some of the best tested results of quantum field theory \cite{zin02,bre10}.
Analogously, utilizing a number of nonperturbative approaches (such as $2+\epsilon$ expansion and numerical evaluations of the path integral), one can extract universal scaling dimensions such as the exponent $\nu$ and the nonperturbative, renormalization group invariant, correlation length scale $\xi$. The latter is in turn related to the vacuum expectation value of curvature, which is measured via a large-scale cosmological constant $\lambda_{obs}$.  
Both the exponent $\nu$ and the scale $\xi$ have significant effects on correlation functions, which deviate from any semi-classical or free field (Gaussian) predictions.  More importantly, these are potentially verifiable via observations, as we will show below. 

\begin{figure} 
	\begin{center}
		\includegraphics[width=0.7\textwidth]{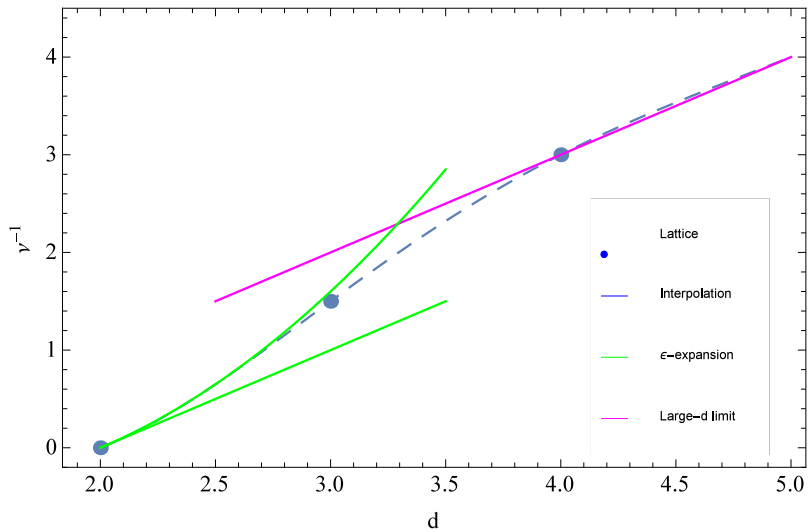}
	\end{center}
	\caption{
	Universal scaling exponent $\nu$
	as a function of spacetime dimension $d$.
    	Shown are the $2+\epsilon$ expansion result to one and two loops \cite{gas78,eps},
	the value in $2+1$ dimensions obtained from the exact solution of the 
       Wheeler-DeWitt equation \cite{htw12}, 
       the numerical result in four spacetime dimensions \cite{ham15}, 
	and the large $d$ result $\nu^{-1} \simeq d-1 \cite{larged,lit04}$.
	}
	\label{fig:exponents}
\end{figure}

In previous work \cite{ham15}, it was shown that for large distances, the invariant scalar curvature correlation fluctuation at fixed geodesic distance behaves as
\begin{equation}
G_R(r) = 
\langle \; \sqrt{g} R(x)  \sqrt{g} R(y) \; 
\delta ( | x - y | -r ) \; \rangle_c \;\; 
\mathrel{\mathop\sim_{r \; \ll \; \xi }} \;\; 
\left( { 1\over r} \right)^{2n}
\;\; .
\label{eq:corr_pow1}
\end{equation}
with a power $2n=2(d-{1 \over \nu})$, where $d$ is the dimension of spacetime.  This shows that the two-point function, for distances smaller than the aforementioned characteristic length scale $\xi$, follows a simple scaling law that is purely dependent on the universal exponent $\nu$.  
The value for the scaling exponent $\nu$ has been calculated and estimated through various means, including $2+\epsilon$ expansion \cite{wei79,gas78,eps}, large $d$ expansion \cite{larged}, numerical lattice calculations \cite{ham15}, exact results in 2+1 dimensions \cite{htw12} and truncated continuum renormalization group methods \cite{reu98,lit04,reu10,cod13,reu14,fal14,fal15,per15,per16,dem14,gie15}.  
Many of these results have been summarized recently, for example, in \cite{ham17}, where it is argued that current evidence points to $\nu \simeq 1/3$ in $d=4$ (see Figure \ref{fig:exponents}).  
In particular, in the large-$d$ limit, one finds $\nu = 1 / (d-1) $ \cite{larged,lit04} in addition to the usual scaling results for the relation for the power in Eq.~(\ref{eq:corr_pow1}), namely $ 2 n = 2 ( d-1/\nu )$.  
This in turn implies for the power $ 2n = 2 $ for $d=4$ and above.  
In the following we will proceed 
on the assumption (supported by extensive numerical calculations on the lattice \cite{ham15}) 
that $\nu =1/3$ exactly in $d=4$, and that consequently the power in Eq.~(\ref{eq:corr_pow1}) is exactly equal, or at least very close to, two.
Another key quantity here is the renormalization group invariant scale $\xi$ appearing for example 
in Eq.~(\ref{eq:corr_pow1}), and related to the quantum gravitational vacuum condensate.
It was argued in \cite{book,ham17,rei14} that the renormalization group invariant $\xi$ is most naturally identified with the scaled cosmological constant via 
$\xi \simeq \sqrt{3 / \lambda}$.  
Modern observational values of $\lambda_\text{obs}$ yield an estimate of $\xi \sim 5300$ Mpc.  
Since galaxies and galaxy clusters involve distance scales of around 
$ r = 1-10$ Mpc $\ll \xi $, the scaling relation in 
Eq.~(\ref{eq:corr_pow1}) should then be applicable for such matter.   
Using these values, one finds in Eq.~(\ref{eq:corr_pow1}) $2n=2(4-3)=2$, and therefore one expects the gravitational curvature fluctuations to scale as
\begin{equation}
G_R(r) = \langle \; \delta R(0) \; \delta R(r) \; \rangle  \sim \; {1 \over r^2} \;\; .
\label{eq:corr_pow1a}
\end{equation} 
Next we proceed to relate the curvature fluctuations $G_R(r)$ to the matter fluctuations $G_\rho(r)$ defined in Section 2.  
As mentioned in the beginning of this section, this procedure is unambiguous because of the gravitational field equations.  
In the limit of matter taking the form of a perfect fluid with negligible pressure
\footnote{This is justified since the clumping of matter, and hence galaxy formation, happens in a matter-dominated era of cosmology, where the perfect fluid and negligible pressure limit is applicable. \label{foot:pressurelessperfectfluid}}, the Ricci scalar satisfies the Einstein trace equation
\beq
R(x) \simeq 8 \pi G \; \rho(x) \;\; .
\label{eq:EFE_trace}
\eeq
Now using its variation
\beq
\delta R(x) \simeq 8\pi G \; \delta \rho(x)
\label{eq:EFE_trace_vary}
\eeq
and inserting this into Eq.~(\ref{eq:corr_pow1a}) gives:
\begin{equation}
\begin{split}
G_R(x) & = (8\pi G)^2 \; \langle \delta \rho(0) \delta \rho(x) \rangle \\
& = (8\pi G)^2 \; \bar\rho^2 \; G_\rho(x) \;\; ,
\end{split}
\label{eq:EFE_trace_vary_1}
\end{equation}
where $G_\rho(x)$ is the matter correlation function defined in Section 2 above. So the scaling behavior of $G_R$ in Eq.~(\ref{eq:corr_pow1a}) unambiguously relates to $G_\rho$ :
\begin{equation}
G_\rho(r) = {1 \over (8\pi G)^2 \; \bar\rho^2} \; G_R(r) \;\; .
\label{eq:GR_scale}
\end{equation}
Since $\bar\rho$ is not expected to scale with distance, the large scale scaling behavior of $G_R$ $\sim r^{-2}$ in Eq.~(\ref{eq:corr_pow1a}) directly gives the $r^{-2}$ scaling behavior of $G_\rho$ in large scales.  Or, in terms of the galaxy correlation index of  Eq.~(\ref{eq:gamma_def}), $\gamma=2$. This is indeed in good agreement with the observed values $\gamma \simeq 1.8$ 
quoted earlier at the end of Sec. \ref{sec:introPk}.
The scaling for the corresponding power spectrum in momentum space can now be determined from the scaling of the real space matter correlation function predicted above.  According to Eq.~(\ref{eq:gamma_3-s}), the prediction of $\gamma=2$ for large scales by quantum gravity implies that $s=1$, or explicitly,
\begin{equation}
P(k)=\frac{a_0}{k}  \; \;  ,
\label{eq:Pk_s=1}
\end{equation}
for small $k$'s.
Note that for large $k$'s (i.e. small distances) however, as the correlation function starts to probe distances far below the average separation of galaxy clusters, or even within one, the standard linear, isotropic assumptions are no longer valid.  In these scales, complex processes and dynamics are expected to dominate over the large-scale gravitational correlation effects that we are considering here.  Hence, for the correlation functions to exhibit a clear $s=1$ scaling, one should consider scales larger than average galaxy cluster separations. 

Typical galaxy clusters have diameters between $1-10 \text{ Mpc}$, and voids, the pockets of empty space between clusters, have typical diameters around $25 \text{ Mpc}$ \cite{dam96}.  
As a result, we would expect the effects of scaling to be most explicit for scales larger than around $50$ Mpc or  $100 \text{ Mpc}$, which correspond to $k$'s smaller than $\lsim 0.15 \; h/\text{Mpc}$.  On the other hand, the spectum is expected to diverge from $s=1$ for small distances (large $k$).
In other words, Eq.~(\ref{eq:Pk_s=1}) implies that for small $k$'s (large scales), the power spectrum should approach a horizontal constant asymptotically on a $k \cdot P(k)$ vs. $k$ plot, with the constant representing the amplitude $a_0$.
In principle, a quantum theory of gravity should also lead to an estimate for the amplitude, as discussed and given in \cite{ham17}.
However amplitudes are generally not universal, and depend on the specific choice of regularization scheme.  
Therefore, in this paper, we prefer to focus on the prediction for the slope, with the amplitude taken as a free parameter that is constrained by fitting to the observational data. 

\begin{figure}
	\begin{center}
		\includegraphics[width=0.75\textwidth]{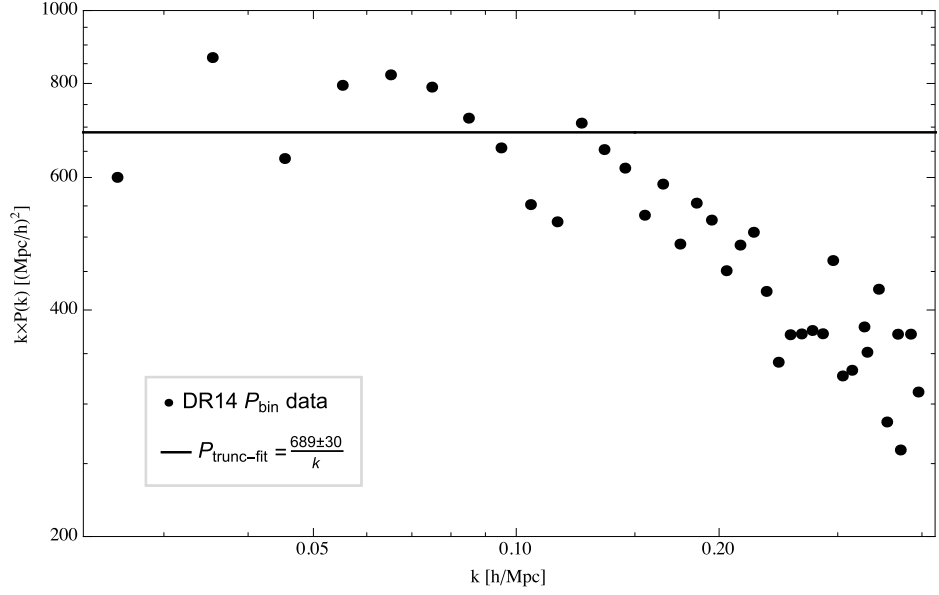}
	\end{center}
	\caption{
	Plot of the observed matter power spectrum versus wavevector $k$.
Black data points are taken from the Sloan Digital Sky Survey (SDSS) collaboration's $14^{\text{th}}$ 2018 data release (DR14) \cite{gil18}.  
The solid line represents the 1-parameter fit for the amplitude $a_0$ to the data, assuming the predicted $s=1$ scaling relation of Eq.~(\ref{eq:Pk_s=1}). 
The vertical axis is plotted with $k \times P(k)$ in order to extract a fitted value of the amplitude $a_0$ in Eq.~(\ref{eq:Pk_s=1}).  
Beyond $k \gsim 0.15 h\text{ Mpc}^{-1}$ the data shows a transient region where the points diverge from the linear scaling, due to the correlation function probing distance scales smaller than the linear scaling regime.
	}
	\label{fig:plot_sdss_pkfit}
\end{figure}

We can now compare the predicted scaling relation of $P(k) = a_0 / k$ with recent galaxy data from the SDSS collaboration measurements on a $k \cdot P(k) \text{ vs. } k$ plot.  
From Figure \ref{fig:plot_sdss_pkfit}, it can be seen that, for sufficiently large distances ($k \lsim 0.15 \; h/ \text{Mpc}$), the general trend of the data indeed converges well on a horizontal line, supporting the $s=1$ prediction from quantum gravity.  
Notice that, as expected, one finds a transient region where the points diverge from the simple linear scaling beyond $k \gsim 0.15 \; h/\text{Mpc}$, due to the correlation function probing distance scales significantly smaller than the linear scaling regime.  
Finally, we can determine the fitted value for the amplitude $a_0$ from the data by the axis intercept on Figure \ref{fig:plot_sdss_pkfit}, giving $a_0 \simeq 689 \pm 30 \text{ (Mpc/\textit{h})}^2$.
One can slightly extend the above analysis by doing a phenomenological fit over the full range of available observational data with two parameters, $a_0$ and $s$, using again the scale-invariant ansatz $ P(k) = a_0 / k^s $, which then gives $s \simeq 1.09$. 
So even if all data points are taken with uniform weight, the quantum gravity prediction of $s=1$ is still consistent with the observational data within a $9\%$ error.
In conclusion, it can be seen that the general trend of the data fits quite well to the $s=1$ slope predicted by quantum gravity, with an amplitude fitted to 
$a_0 \simeq 689 \pm 30 (Mpc / h )^2$. 
We note here that the amplitude $a_0$ is, in principle, calculable from the lattice treatment as discussed for example in \cite{ham17}, nevertheless it represents a non-universal quantity and 
thus clearly depends on the  specific way the ultraviolet cutoff is implemented in the quantum theory
of gravity, a fact that is already well known in lattice QCD \cite{das81}.
Here we choose instead to take the non-universal amplitude $a_0$ as an adjustable
parameter, to be fitted to the observational data.

\vskip 20pt

\section{Transfer Function and Normalization of the Power Spectrum} \label{sec:Tk}   
\vskip 10pt

So far, our analysis asserted that the scaling of galaxy distributions, which is governed by that of matter fluctuation correlations $G_\rho (r)$, is directly calculable from curvature fluctuation correlation functions $G_R (r)$ in the quantum theory of gravity.  
In particular, the matter and curvature fluctuations are unambiguously connected via the Einstein field equations.  
Usually there is an implicit assumption that the $T_{\mu\nu}$ on the right hand side takes the perfect fluid form, which involve an equation of state of pressure-less matter.  
The latter assumption is certainly valid for the matter-dominated universe
\footnote{see footnote \ref{foot:pressurelessperfectfluid} in Sec. 3}, 
an era in which we expect galaxy-formations to take place.  
As can be seen in Figure \ref{fig:plot_sdss_pkfit}, the observational data is in rather good agreement with the s=1 prediction of quantum gravity for the scale of galaxy clusters 
(i.e. $ k \lesssim 0.15 \text{ hMpc}^{-1}$).  

In principle, quantum gravity with its long-range correlations and $n$-point functions should also govern and make predictions beyond galactic observation scales.  
Below galactic scales (large $k$), complex and non-linear interactions between matter arises, and gravitational correlations are expected to become subdominant effects and not easily discernible.  
On the other hand, beyond the scale associated with galaxy clusters and superclusters (small $k$), gravity is again undoubtedly the dominant long-range force.  
There we expect fluctuations in curvature to again play a dominant role in governing long-range correlations.  
However, observationally, larger spatial scale corresponds to looking at increasingly earlier epochs of the universe, perhaps even before galaxies are formed.  
Therefore data from galaxy surveys cannot serve as a test to the predictions in these regimes.  
Fortunately, modern astronomy has provided very detailed measurements of the CMB, 
which encodes fluctuations over a wide range of cosmological scales.  
This can then be used as an additional test for the quantum gravity predictions.  

In this section, we will discuss how we can utilize the so-called transfer function to extrapolate the quantum gravity prediction to increasingly small $k$ far beyond galaxy scales, 
and compare to how it fits observed CMB data. 
It should be noted that in the current CMB literature one often refers to the angular power spectrum $C_l$ as the primary quantities obtained.  
Nevertheless, $P(k)$ is directly related to $C_l$, as described in many standard 
literature works on the subject \cite{wei08,lid00,dod03,pee93}.  
However here instead we focus on $P(k)$, since, as mentioned above, it is this quantity whose scaling is most directly related to that of curvature fluctuations, the quantity that is predicted from quantum gravity. 
According to the Friedman equations, wavelengths that enter the horizon before matter-radiation equality evolve differently than those that enter after.  
As a result, as we look at even larger scales (smaller $k$'s), the $s=1$ slope discussed previously is expected to change.  
Nevertheless, we can use a well-known function in cosmology, the so-called transfer function $\mathcal{T}(\kappa)$, to relate small wavelength behaviors to large wavelength behaviors.  
Details of which can be found in standard texts such as \cite{wei08}; for explicit numerical results 
see also \cite{whu97}.
Here we will outline below how the use of the transfer function can be applied to our predictions. 

We recall the definition of the spectral function 
\begin{equation}
\label{eq:Pk_def2} 
P(k) = (2\pi)^3 F(t_0 )^2 \langle \; |\Delta(k)|^2 \; \rangle \;\; . 
\end{equation} 
Here $F(t)$ can be calculated from the Friedman equations, giving 
\begin{equation}
\label{eq:Ft} 
F(t)=\frac{3}{5} \left [ {a(t) \over a_L} \right ] C(x)  \;\; , 
\end{equation} 
where $x=x(t)=(\Omega_\Lambda / \Omega_M)  \, a(t)^3 $ , and the correction factor $C(x)$ is given by 
\begin{equation}
\label{eq:Cx} 
C(x) \equiv \frac{5}{6} \; x(t)^{-\frac{5}{6}} \sqrt{1+x(t)} 
\int_{0}^{x(t)} du \, {1 \over u^{\frac{1}{6}} (1+u)^{\frac{3}{2}} } \;\; . 
\end{equation} 
Detailed steps in the above derivations can be found in \cite{wei08}.  
As for the function $\Delta (k)$ in (\ref{eq:Pk_def2}), it can be calculated from the Boltzmann transport equations, again presented in detail in \cite{wei08} and references therein, giving 
\begin{equation}
\label{eq:Delatq} 
\Delta(q) \approx \delta(q,t_L) - t_L \; \psi(q,t_L) 
= \frac{2}{3} \cdot \frac{q^2 \, \mathcal{R}_q^0 \; \mathcal{T}(\kappa)}{H_L^2 a_L^2} \;\; , 
\end{equation}
where the comoving wavenumber $q$ is often used in place of the physical wavenumber $k$, the two being related by $k \equiv q/a(t)$.  
Here $\delta(q,t)$ is the matter density contrast defined in Eq.~(\ref{eq:delta_def}) in momentum space, $\psi(q,t)$ is the gravitational perturbation in Newtonian gauge, $t_L$ is the time of decoupling of radiation from matter, associated with the recombination of hydrogen, and $H_L$  and $a_L$ are the Hubble rate and scale factor correspondingly evaluated at $t_L$.  
The last equality is obtained by conveniently parameterizing the combination of $\delta$ and $\psi$ with $\mathcal{R}_q$ and $\mathcal{T}(\kappa)$, known as the comoving curvature perturbation and transfer function, respectively \cite{wei08}.  
The transfer function $\mathcal{T}(k)$ is usually expressed with an argument $\kappa \equiv \sqrt{2} \left(  k / k_{EQ} \right) $ , where $k_{EQ}$ is the scale at matter-radiation equality.
Inserting these results into the definition for $P(k)$ in Eq.~(\ref{eq:Pk_def2}) then gives 
\begin{equation}
P(k)= \frac{4(2\pi)^2 a_0^3 \;  C^2(\frac{\Omega_\Lambda}{\Omega_M})}{25 \; \Omega_M^2 H_0^4} \;  {\mathcal{R}_q^0}^2 k^4 \; \left[ \mathcal{T} \left( \sqrt{2} \frac{k}{k_{EQ}} \right) \right]^2 \;\; .
\label{eq:Pk_full}
\end{equation}
Again, following common convention \cite{wei08}, the function $\mathcal{R}_q$ is often parametrized as a simple power of $q$
\begin{equation}
\mathcal{R}_q \simeq N q^{-3/2} \;  {\left( \frac{q}{q_*} \right) }^{(n_s-1)/2} ,
\label{eq:Rq_param}
\end{equation}
where $q_*$ is some arbitrarily chosen reference scale, and $n_s$ is referred to as the spectral index.
Then $P(k)$ can be conveniently factorized into
\begin{equation}
P(k) = C_0 \; A \; k^{n_s} \left[ \mathcal{T} (\kappa) \right]^2  ,
\label{eq:Pk_fullfact}
\end{equation}
with 
\begin{equation}
A \equiv \frac{N^2 }{ k_*^{n_s-1}  }  
\text{ , and } 
C_0 \equiv \frac {4 \, (2\pi)^2  
\left[ C \left( \frac{\Omega_\Lambda}{\Omega_M} \right) \right]^2}  
{ 25 \, \Omega_M^2 H_0^4    } \; \; ,
\end{equation}
so that $C_0$ is a prefactor that encodes exclusively cosmological model parameters.
It was Harrison and Zel'dovich who originally suggested in the seventies that $n_s$ should be close to one \cite{hz70}.

It is known that the transfer function $\mathcal{T}(\kappa)$ is entirely determined from classical cosmological evolution and by the solutions of the associated coupled Boltzmann transport 
equations for matter and radiation.
It will turn out to be useful here that the function in question can be accurately described 
by the semi-analytical formula given explicitly by D.~Dicas as quoted in \cite{wei08,whu97},
\begin{equation}
\mathcal{T} (\kappa) \simeq \frac{\ln[1+(0.124\kappa)^2]}{(0.124\kappa)^2} 
\left[  
{ {1+(1.257\kappa)^2+(0.4452\kappa)^4+(0.2197\kappa)^6} 
\over  
{1+(1.606\kappa)^2+(0.8568\kappa)^4+(0.3927\kappa)^6} }  
\right]^{1/2} \; \; ,
\label{eq:Tk}
\end{equation}
and which will be used here in the following discussion.
For later reference, the shape of the function $\kappa \vert \mathcal{T} (\kappa) \vert^2 $ is shown below in Figure \ref{fig:plot_kTk} ; 
particularly noteworthy here is the inverted-v shape reflecting the cosmological 
evolution transition from a radiation dominated to a matter dominated universe, 
again as discussed in great detail in \cite{wei08}.

Now, both $C_0$ and $\mathcal{T}(\kappa)$ are fully determined, the former by cosmological measured parameters from, for example, the latest Plank satellite data \cite{ade15,mac15}, and the latter theoretically from the Boltzmann transport equations, numerically evaluated in \cite{wei08,whu97} and further references cited therein.
Therefore the end result is that $P(k)$ is essentially parameterized by two quantities,  an overall amplitude $A$ and the spectral index $n_s$.
\begin{figure} 
	\begin{center}
		\includegraphics[width=0.8\textwidth]{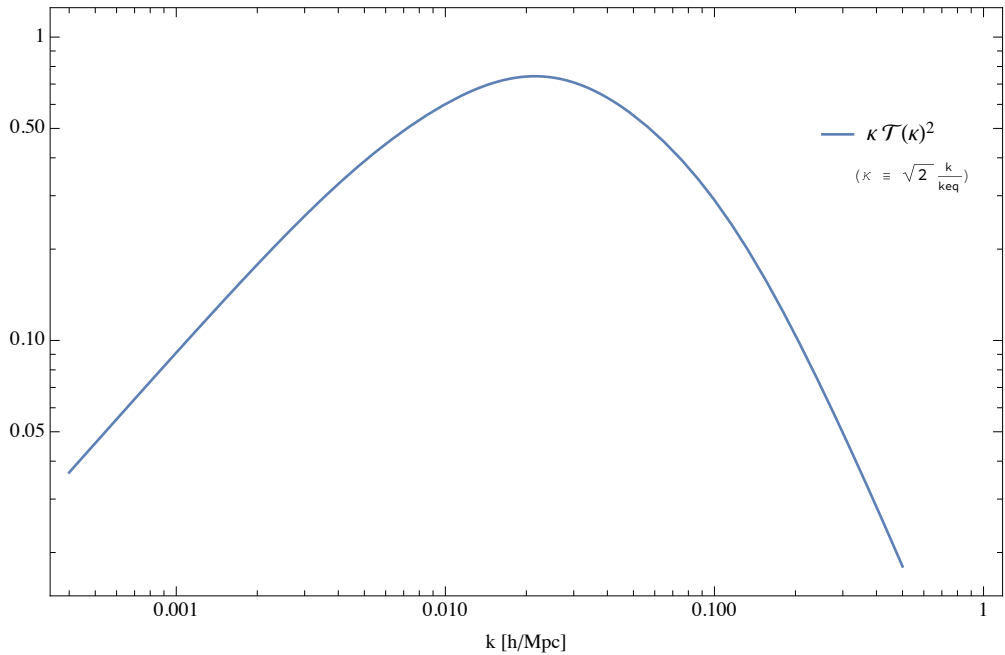}
	\end{center}
	\caption{
Shape of the transfer function $\kappa \, \vert \mathcal{T} (\kappa) \vert^2 $ using the interpolating formula of Eq.~(\ref{eq:Tk}).
		What is of interest here is the significant turnover happening at wavenumbers 
       $k \sim 0.02$, which is known to correspond to a cosmological time scale 
       associated with matter-radiation equilibrium.
This marked turnover here is the primary reason for the peculiar inverted-v shape of the power spectrum in the following plots.}
\label{fig:plot_kTk}
\end{figure} 
The next step is to use quantum gravity to theoretically constrain the value of $n_s$, and analyze how well a power spectrum with the predicted value of $n_s$ fits the current observational data.
As discussed previously, quantum gravity predicts $ P(k) \sim a_0 / k^s $  with an exponent $s=1$ in the galaxy regime (i.e. $ k \sim 0.01 \text{ to } 0.3 \text{ }h\text{Mpc}^{-1}$).  
In the previous section, we noted that although $a_0$ is in principle calculable, additional subtleties arise.  
Hence, for the current purpose, we simply use the value that fits well the galaxy data at the largest scales, namely $a_0 \simeq 689$.  
As a result, within the galaxy regime, we should have a matching
\begin{equation}
P(k) \equiv C_0 \; A \; k^{n_s}  \, \mathcal{T}(\kappa) \; 
\approx {689 \over k} \; \; \; \;
(\text{ for } k \sim 0.01 \text{ to } 0.3 \text{ }h\text{Mpc}^{-1})  \;\; .
\label{eq:Pk_equate}
\end{equation}
As mentioned earlier under Eq.~(\ref{eq:Tk}), both $C_0$ and $\mathcal{T}(\kappa)$ are fully determined from the classical cosmological FRW evolution 
equations, leaving this essentially an equality with two unknowns, namely $A$ and $n_s$, that holds in the galaxy regime.  
Therefore, by appropriately selecting two points from the quantum gravity prediction within this regime, we can derive the values for $A$ and $n_s$.  
Since the left hand side is supposedly valid for all scales, with the two values $A$ and $n_s$ determined, these will then give us a power spectrum that allows us to extrapolate the quantum gravity prediction within the galaxy regime to much larger scales.

Within the galaxy regime, there is a certain flexibility in which two points one may choose.  
For the purpose of a first estimate, we select two points relatively apart, but not too close to the margins of the available galaxy data.  
Although the data set ranges from $0.025 \lsim k \lsim 0.3 \; h\text{ Mpc}^{-1}$, the linear scaling regime is only valid up to around $k \lsim 0.15 \; h\text{ Mpc}^{-1}$, 
as discussed in Sec. \ref{sec:QGexpofPk}. 
On the other hand, extreme small $k$'s (large distances) are expected to be less representative, as they suffer from limited statistical samples within the ensemble.	 
Thus, an appropriate preliminary choice, for first-estimation purpose, would be:
\begin{equation}
\begin{split}
k_1 &= 0.035 \; h\text{ Mpc}^{-1} \\
k_2 &= 0.10 \; h\text{ Mpc}^{-1}  \; \; .
\end{split}
\label{eq:k1k2}
\end{equation}
These correspond to real space scales of
\begin{equation}
\begin{split}
\lambda_1 &\sim 250 \text{ Mpc   }\\
\lambda_2 &\sim 90 \text{ Mpc} \; \; .
\end{split}
\label{eq:l1l2}
\end{equation} 
Using these test points in Eq.~(\ref{eq:Pk_equate}) gives
\begin{equation}
n_s = 1.108 \;\; , \;\; A=5.97 \times 10^{11} \;\; .
\label{eq:ns_A}
\end{equation}
The best fit from the Plank 2015 data gives a fit of $n_s=0.9667 \pm 0.0040$ \cite{ade15,mac15}.  
This suggests our preliminary analysis, using the two above selected points, yields a value of $n_s$ that is around $\sim 15\%$ of the best fit value of the data, as shown in Figure 4. 

There are two flexible aspects in this analysis.  
Firstly, there is a flexibility in choosing $k_1$ and $k_2$, and secondly there are intrinsic statistical errors in the observational data.  
As an example, by purely adjusting the former, it is possible to get a value of $n_s \approx 0.9827$ if we use 
$ k_1  \sim 0.040 \; h \; \text{Mpc}^{-1} $ and $ k_2  \sim 0.065\; h \; \text{Mpc}^{-1} $, 
which correspond to 
$ \lambda_1 \sim 230 $ Mpc and $ \lambda_2 \sim 140 $ Mpc, respectively.  
Though this normalization choice seem to yield a value for $n_s$ closer to that of Plank, the two points do not seem sufficiently apart to normalize the spectrum properly in our quantum gravity picture.  Nevertheless, this methodology also provides a rough estimate for the overall uncertainty in $n_s$.
\begin{figure} 
	\begin{center}
		\includegraphics[width=0.8\textwidth]{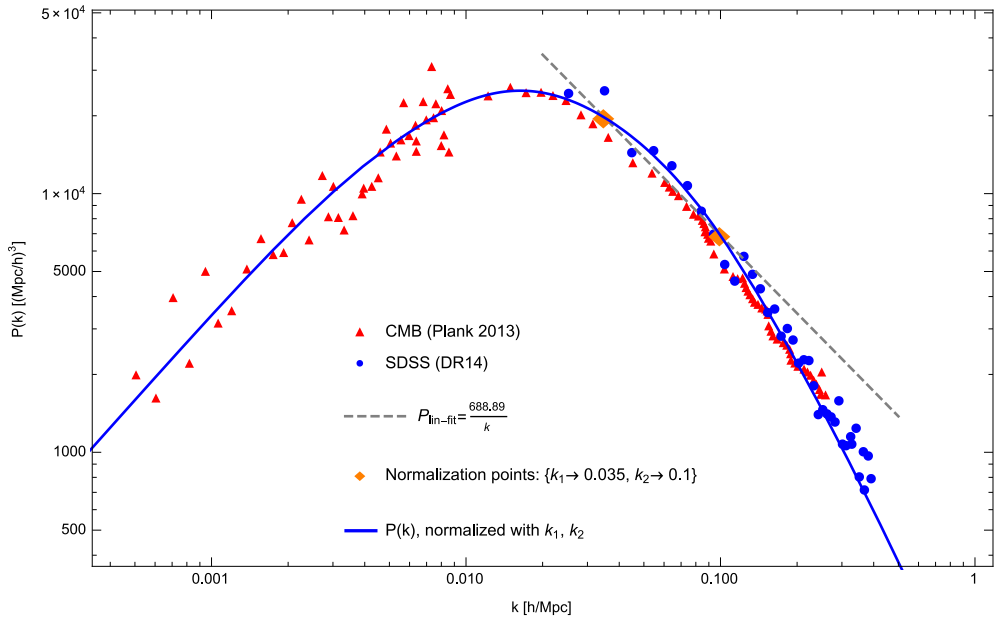}
	\end{center}
	\caption{
Full Power Spectral Function $P(k)$ (blue solid) normalized by $P_{lin}=a_0/k^s$ with $s=1$, $a_0=688.9$ (gray dashed), and two reasonable normalization points (orange).  This choice of normalization yields a value of $n_s=1.108$ and $A=5.97 \times 10^{11}$, which fits well with both CMB (red) and SDSS (black) data. }
\label{fig:plot_pkcomb}
\end{figure}

\vskip 30pt

\section{Comparison with Inflationary Models}
\label{sec:compare}

\vskip 10pt

A popular attempt to explain the matter power spectrum is through inflation models, which proposes that the structures visible in the Universe today are formed through fluctuations in a hypothetical primordial scalar inflaton field.  Although the precise behavior and dynamics of the inflaton field are still largely controversial \cite{ste11,ste12,teg05,ste13,ste14,ste17}, there are some general features common among these models that allow one to derive predictions about the shape of the power spectrum.  Here we very briefly outline the inflation perspective, and how it radically differs from the quantum gravity perspective we have presented earlier.  
For more details and reviews on the inflation mechanism, we refer to recent literatures such as the books \cite{wei08}, \cite{lid00} and \cite{dod03}.

Cosmic inflation was first proposed in \cite{gut81,lin82,alb82} as a possible solution to the horizon and flatness problem in standard big bang cosmology, by proposing a period of exponential expansion in the early phases of the universe. This exponential expansion is expected to be driven by a hypothetical``inflaton field'', usually scalar in nature, which dominates the energy density in the early universe, causing a de Sitter type accelerated expansion.
In these models, the gravitational perturbations that ultimately cause gravitational collapse and structure formations are also due to fluctuations in this inflaton field.  Scalar field fluctuations are assumed to be Gaussian for large enough distances (small $k$), which then naturally results in a value for the spectral index $n_s=1$.  
More realistic models with non-zero tensor to scalar ratio predicts $n_s$ between $0.92$ and $0.98$ \cite{teg05,ste04,ste06,gut14}, without assuming excessive fine-tuning of parameters \cite{teg05}
\footnote{
The measured value of $n_s$ is often used to back infer the inflation tensor to scalar ratio, $r$, which gives around $r \lesssim 0.11$ for $n_s=0.968 \pm 0.006$ \cite{ade15,mac15}.}
The above value of $n_s=1$ is then used in Eqs.~(\ref{eq:Pk_fullfact}) and (\ref{eq:Tk}), leaving only one free parameter, the amplitude $A$ in Eq.~(\ref{eq:Pk_fullfact}), which can then be determined by normalizing to the CMB data at very large scales (i.e. small $k$).

Notice that the gravitational perspective we suggest in this paper differs fundamentally from that of inflation, both in procedure and, most importantly, in origin.  
Firstly, inflation models suggest that gravitational perturbations are due to fluctuations in an inflaton field, whereas here perturbations are intrinsically gravitational and quantum mechanical in origin.  Secondly, inflation models use the corresponding inflaton correlation function, $G_\phi (r)$, or power spectrum, $P_\phi (k)$, which are mostly scalar/Gaussian in nature, to derive those of matter $G_\rho (r),P(k)$, whereas in our
gravitational picture, the correlation function that constrains $G_\rho (r)$ and $P(k)$ is the curvature $G_R (r)$, which is highly non-Gaussian.  
Finally, while both perspectives provide a prediction for $n_s$ which governs the shape of $P(k)$, both leave the overall normalization $A$ in Eq.~(\ref{eq:Pk_fullfact}) uncertain. 

Modern renormalization group theory would imply that the universal critical exponent $\nu$, and the scaling dimensions $n$ that follow from it [see Eq.~(\ref{eq:corr_pow1})], are expected to be {\it universal}, and as such only dependent on the spacetime dimension, the overall symmetry group, and the spin of the particle.
The same cannot be said of the critical amplitudes, such as the amplitude associated with the curvature two-point function of Eq.~(\ref{eq:corr_pow1}).
In principle it is possible to estimate from fist principles the amplitude $A$ and thus the normalization constant ${N}$ is Eq.~(\ref{eq:Rq_param}), since these can be regarded as wave function normalization constants in the underlying lattice theory.  
A rough estimate was given in \cite{ham17}, nevertheless a number of reasonable assumptions had to be made there in order to relate correlations of small gravitational Wilson loops to very large (macroscopic) ones. 
As a result, the estimate for the amplitude quoted there is still two orders of magnitude smaller than the observed one.
Nevertheless it is understood that amplitudes are expected to be regularization and scheme dependent, and will differ by some amount depending on the specific form of the ultraviolet cutoff scheme chosen, whether it is a lattice one, or a continuum inspired momentum cutoff, or dimensional regularization.
This phenomenon is well known in scalar field theory as well as lattice gauge theories, and allows the relative correction factors to be computed to leading order in perturbation theory \cite{das81}, thus reducing the regularization scheme dependent uncertainties. 
In the following we choose to leave, for now, the overall amplitude in Eqs.~(\ref{eq:corr_pow1}) or (\ref{eq:Pk_s=1}) as a free parameter, and constrain it instead directly by the use of observational data.

Inflation models on the other hand often normalize the curve at large scales with CMB data or at the turnover point $P_\text{max}$  \cite{wei08}, but our picture normalizes it with the slope within the galaxy domain, where we argue that the curvature-matter relation is most direct, 
as discussed in Sec. \ref{sec:QGexpofPk}.

We contend that there are a number of reasons that the gravitation-induced picture is more natural for matter distributions than the inflaton-induced one.  
Firstly, the gravitational field is a well-established interaction and is supposed to have long-range influences with cosmological consequences, whereas the scalar inflaton field remains an observationally unconfirmed quantum field whose precise properties, such as its potential and interaction with standard model fields, remain unknown.  
Secondly, the Feynman path integral treatment is an unambiguous procedure to quantize a theory covariantly, producing in principle a unique theory with essentially no free parameters (as in the case of QED and QCD).  

On the other hand, there is little consensus on the precise dynamics of the inflation field, the number of competing models is far from unique, and almost all suffer from fine-tuning problems, leading to ad-hoc potentials.  
Many also question the falsifiability of such theories.  
Physicists well known cosmologists, including some of the creators of the theory, have expressed various issues with this paradigm \cite{ste11}.  
Therefore, whether or not inflation theories are a satisfactory solution to the cosmological problems are still in question.

At this stage, despite the differences in perspective, there is yet no clear-cut preference over each other except for the naturalness arguments given above.  
However, a number of second order effects due to quantum gravitational fluctuations could yield diverging and distinct predictions, including the IR regulation from $\xi$, the RG running of G, 
and n-point functions, which should be testable in the near future with increasingly accurate measurements.  
So in contrast, the predictions from quantum gravitation are (i) unique and (ii) falsifiable.  Therefore, we believe this provides a compelling alternative to inflation.  We will now discuss these new effects in the next section.

\vskip 20pt

\section{Additional Quantum Effects in the Small-$k$ Regime}
\label{sec:qmeffects}

\vskip 10pt

So far we have introduced the alternative picture that matter perturbations are directly resulted from, and hence governed by, underlying quantum fluctuations from gravity, instead of the inflaton.  
In this section we discuss two additional quantum-mechanical effects as a result of gravitational fluctuations, which would be explicitly distinct from any scalar field inflationary models.  
These are (i) the presence of an infrared (IR) regulator, and (ii) the renormalization group (RG) running of Newton's constant $G$. 

In a quantum field treatment of gravity, a nonperturbative scale $\xi$ is dynamically generated which regulates the otherwise serious infrared divergences associated with a zero mass graviton.  
In addition, Newton's constant $G$ is seen to ``run'' with scale as a result of quantum corrections, in analogy to what happens in QED and QCD.  
In momentum space, the formula for the running of $G$ is given in \cite{hw05,ham17} by

\begin{equation}
G(k) \; \simeq \; G_0 \left[ 1+ c_0 \left( \frac{m^2}{k^2} \right)^{3/2} 
\, + \, \mathcal{O} ( \left(\frac{m^2}{k^2} \right)^3 ) \right] \; .
\label{eq:runG}
\end{equation}
Here $G_0$ is the classical (i.e. $k \gg m$) ``laboratory'' value of the Newton's constant, or simply $G$ henceforth in the paper, $m$ is the renormalization scale in momentum space related to $\xi$ by $m\equiv 1/\xi$, and $c_0$ the coefficient for the overall amplitude of the quantum correction, which is generally expected to be of order one.  
As for the nonperturbative scale $\xi$, it is most naturally related to the cosmological constant $\lambda_0$ \cite{book,rei14} by
\begin{equation}
m^2 \equiv \frac{1}{\xi^2} \simeq \frac{\lambda_\text{obs} }{3}  \; \; .
\label{eq:m^2Toxi^2}
\end{equation}
Notice that these two quantum effects are not unique to gravity, but common in quantum field theories.  The latter is most representative in the well-studied theories of QED and QCD, where quantum corrections result in the running of coupling constants $e$ and $\alpha_s$ respectively.  
It is well known that a dynamically generated infrared cutoff arises in other nonperturbative theories such as the nonlinear sigma model, QCD and generally non-Abelian gauge theories.  
The scale $\xi$ therefore plays a role analogous to the scaling violation parameter $\Lambda_{\overline{MS}}$ in QCD.  $\xi$ also serves directly as a characteristic scale in the theory that, much like the $\Lambda_{\overline{MS}}$ scale in QCD, distinguishes the small distance from the large distance domain.  
Both effects are discussed in detail in \cite{book,ham17}.

The formal implementation of the dynamically generated scale $\xi$ is to be inserted as a lower infrared cutoff in any momentum integrals.  
This would mean amending 
\begin{equation}
\int_{-\infty}^{} dk \; \; \rightarrow \; \; \int_{m} dk  \; \; .
\end{equation}
However, a more straightforward implementation is to simply make replacements $k^2 \rightarrow k^2+m^2$ with $m=1/\xi$, which phenomenologically works well in other nonperturbative theories (such as QCD) and partially includes the effects of infrared renormalons \cite{ben99}.
As a result, for the power spectrum given in Eq.~(\ref{eq:Pk_fullfact}) one obtains
\begin{equation}
\begin{split}
 P(k) & = C_0 \cdot A \cdot k^{n_s}  \left[ \mathcal{T} ( \sqrt{2} \;  {k \over k_{EQ}} ) \right]^2 \; \\
\rightarrow \; \; 
 P(k)_\text{reg} & = C_0 \; A \cdot (k^2+m^2 )^{\frac{n_s}{2}} \; \left[ \mathcal{T} \left( \sqrt{2} \,  {(k^2+m^2 )^{1/2} \over k_{EQ}} \right) \right]^2 \;\; .
\end{split}
\label{eq:Implement_IRreg}
\end{equation}
The effect of this modification can be seen as the orange curve in Figure \ref{fig:plot_pkregrun} below.  It is most visible in the extreme long distance (small $k$) domain where $k \sim 1/ \xi$.

Finally, another important expected quantum effect is the renormalization group running of Newton's $G$, which was already discussed in some detail in \cite{hw05,book,rei14} and implemented either through a scale dependence in momentum space $G(k)$ or by the use of a set of covariant nonlocal effective field equations containing a $G (\Box)$, where 
$\Box \equiv g^{\mu\nu} \nabla_\mu \nabla_\nu $ is the covariant d'Alembertian 
acting on $n$-th rank tensors.
Here, to implement the running of Newton's $G$, one needs to identify where $G$ would appear in the power spectrum. 
Recall the matter correlations are related to curvature correlations via the Einstein field equations
\begin{equation}
G_\rho (x)= \frac{1}{(8\pi G)^2}  \; (\bar{\rho})^{-2} \, G_R (x) \; ,
\label{eq:G_rhoToG_R}
\end{equation}
or in momentum space, via a Fourier transform
\begin{equation}
P(k) =  \frac{1}{(8\pi G)^2}  \; (\bar{\rho})^{-2} \, P_R (k) \; ,  
\label{eq:Pk-PkR}
\end{equation}
where, as usual, $P(k)$ is the spectrum for matter fluctuations, and 
$P_R (k) \equiv \, <|\delta R(k)|^2> \sim 1/k^s$ with index 
$s=1$ is the spectrum for curvature fluctuations. 
Observing that $P(k) \propto 1/G^2$ dimensionally, promoting $G \rightarrow G(k)$ can be achieved by the replacement
\begin{equation}
	P(k) \rightarrow \; \;  P(k)_{\text{run}} = \left[ \frac{G}{G(k)} \right]^2 P(k)
\end{equation}
with $G(k)$ given in Eq.~(\ref{eq:runG}).
In addition, since the running of $G$ is again an effect that is most significant on small $k$'s, the strong infrared divergence near $k \simeq 0$ should similarly be regulated as discussed above correspondingly, by the replacement $k^2  \rightarrow k^2  + m^2$ in $G(k)$.  
As a consequence, the fully IR regulated expression with the running factor is
\begin{equation}
P(k)_\text{run} = \left[ 1 + c_0 \left( {m^2 \over k^2+m^2} \right)^{\frac{3}{2}} \right]^{-2} 
\cdot 
C_0 \; A \left(k^2+m^2\right)^{\frac{n_s}{2}} 
\cdot
\left[ \mathcal{T} \left( \sqrt{2} \;  {(k^2+m^2 )^{1/2} \over k_{EQ}} \right) \right]^2
\label{eq:Implement_RGrun}
\end{equation}
This expression is plotted as the green curve in Figure \ref{fig:plot_pkregrun}.
\begin{figure}
	\begin{center}
		\includegraphics[width=0.8\textwidth]{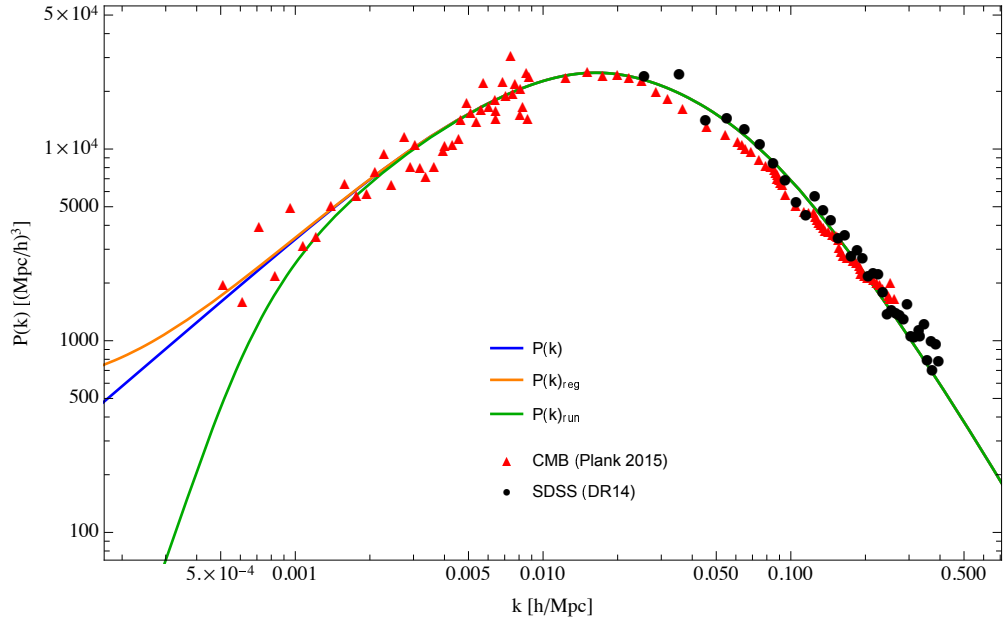}
	\end{center}
	\caption{
		Effects of the IR regulator and RG running of G.  
                             The red triangles here show the 2015 Planck CMB data \cite{ade15,mac15},
                             while the black circles refer to the latest Sloan Digital Sky Survey (SDSS) 2018 
                             data release \cite{gil18}.  
       The blue curve represents the full power spectral function $P(k)$ as derived 
       and shown previously in Figure \ref{fig:plot_pkcomb}.  
       The orange curve $P(k)_{\text{reg}}$ includes the effect of an IR regulator,  
       while the green curve $P(k)_{\text{run}}$ includes the effects of an IR regulator
       together with the RG running of Newton's G.
		}
	\label{fig:plot_pkregrun}
\end{figure}

\begin{figure}
	\begin{center}
		\includegraphics[width=0.8\textwidth]{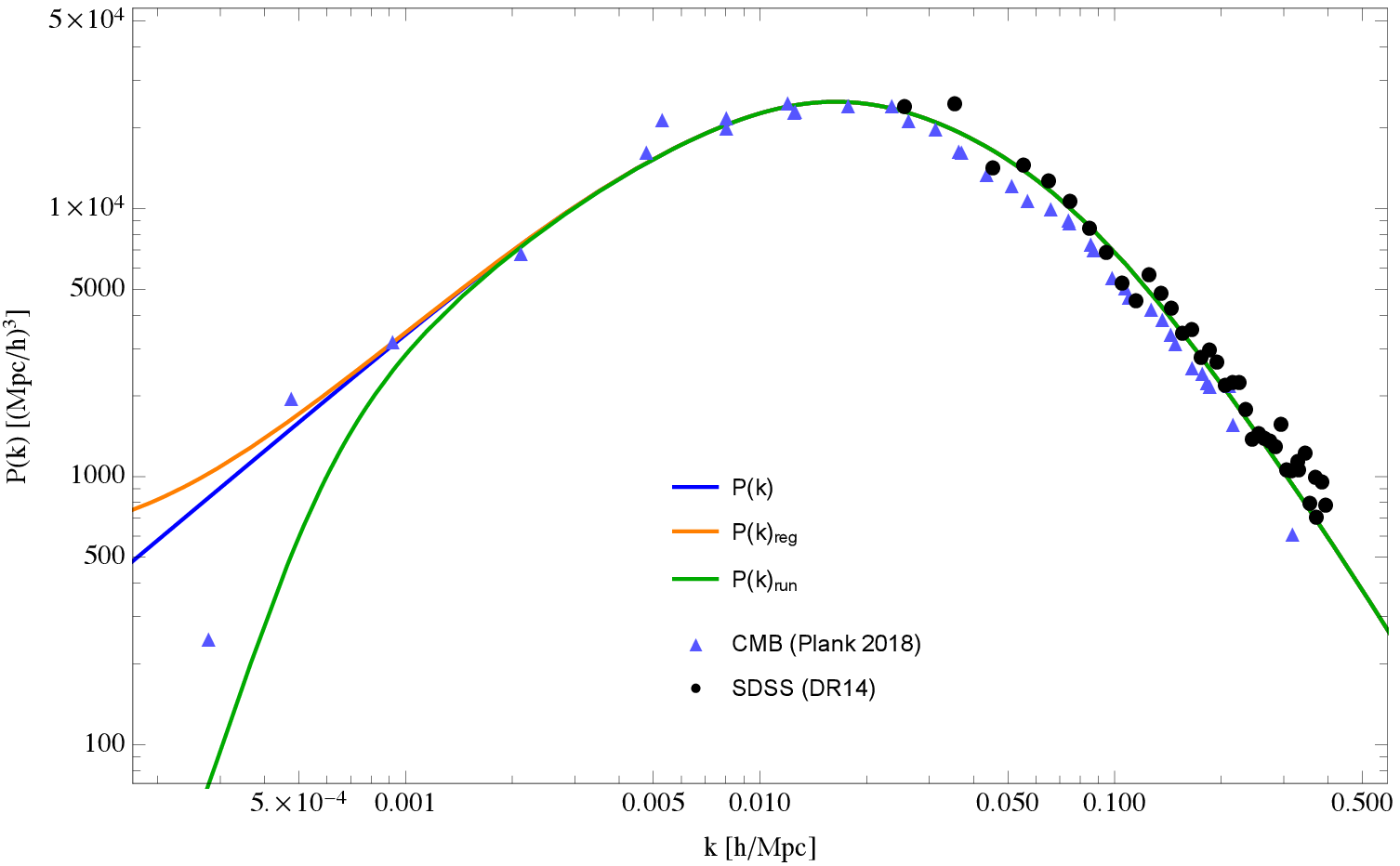}
	\end{center}
	\caption{
Same plot as in Figure 5, but now using the latest July 2018 Planck CMB data release                                              \cite{akr18} (blue triangles), showing again the effect of the IR regulator and of the RG running of $G$.  
As before, the blue curve represents the full power spectral function $P(k)$ as derived 
       and shown previously in Figure \ref{fig:plot_pkcomb}, 
       the orange curve $P(k)_{\text{reg}}$ includes the effect of an IR regulator, 
       and the green curve $P(k)_{\text{run}}$ includes the effects of an IR regulator 
       together with the RG running of Newton's $G$.
       Notice that the new observational data set includes additional 
       low-$k$ data points, which seem
       to support (within large errors for the new data points) the lowest 
       $P(k)$ (green) curve, 
       implemented here with the IR regulator and the RG running of $G$.
       	}
	\label{fig:plot_pkregrun1}
\end{figure}

In summary, from Figures \ref{fig:plot_pkregrun} and \ref{fig:plot_pkregrun1} one can see that the effects of the IR regulator alone serves to level off the curve at a low $k \sim m=1/\xi$ (orange curve), whereas the full modification (green curve), which includes both the IR regulator and the RG running of $G$, bends the curve downwards even further (blue curve).  
Indeed the newest CMB observational data \cite{akr18} shown in Figure \ref{fig:plot_pkregrun1}, now including an additional data point at an even lower value of $k \sim 2.8 \times 10^{-4} h Mpc^{-1} $, actually suggests a downwards bend in the low-$k$ regime, deviating from the classical prediction (blue curve), and approaching the quantum prediction (green curve).  
Although the error bars on the leftmost data point still remains rather large, it is somewhat auspicious that it expresses such a downwards bend.  
One might therefore be hopeful that as the observational resolution continues to improve, 
error bars for the data in the low-$k$ regime will continue to narrow down, 
eventually confirming or falsifying this important prediction.

It should be reiterated that the quantum gravity prediction of the IR dynamical regulator and of the running of $G$ (shown by the green curve) is independent on whether the origin of the matter fluctuations is inflation-driven or gravity-governed.  
Thus, even if one insists the power spectrum must be inflation-driven, or some sort of combined effect between inflation and gravitation, the data suggests that the IR regulation and RG running of $G$ are quantum gravitational effects that should be taken into account.  
Or, at the very least, these nonperturbative quantum effects due to gravity form a rather compelling explanation for this observed dip in the low-$k$ region.

Once again, it should be stressed that these two additional, genuinely quantum, effects are fairly concrete predictions associated with quantum modifications to classical gravity, and are quite distinct from any inflationary models, both in procedure and in origin.  
With the advance of increasingly accurate astrophysical and cosmological satellite observations, it is hoped that these new predictions could be verified, or falsified, in the near future.

\vskip 20pt

\section{Conclusion}
\label{sec:conclusion}

\vskip 10pt

In this paper we have provided an explanation for the galaxy and cosmological matter power spectrum that is purely gravitational in origin, to our knowledge first of its kind without invoking inflation.  We first showed how gravitational fluctuations would unambiguously govern the slope of the spectrum in the galaxy survey domain.  As seen in Figure \ref{fig:plot_sdss_pkfit}, the prediction is in very good agreement with observations.  Next, using the transfer function we extrapolated the prediction to all scales, which is also in general agreement with observations (Figure \ref{fig:plot_pkcomb}).  Later, we investigated additional quantum effects that are expected to be important at large scales (small $k$), and show how these predictions diverge from conventional inflation-inspired predictions (Figure \ref{fig:plot_pkregrun}).

The primary benefit of this explanation over inflation is that there is no need for additional hypothetical ingredients of physics, other than Einstein's gravity and accepted quantum field theory methods.  The correlation functions of gravitational fluctuations are fully governed by correlators in a quantum field theory of gravity, as for fluctuations of any other interaction.  
It is widely known that gravity is nonrenormalizable in a perturbative treatment; hence standard nonperturbative techniques based on the Wilson-Parisi renormalization group analysis \cite{wil72,par73,par76} need to be used, and then the respective critical exponents are fully calculable. 
In this paper we show how these results translate to predictions for relative spectral indices for matter fluctuations unambiguously via the field equations.  
It is argued that the further extension of this result to all scales only requires the transfer function, a well-known ingredient in cosmology that involves mainly collision analysis via the Boltzman equations.  
In this picture, no new physical ingredients are postulated, nor necessary.  
On the other hand for inflation, a new, minimum of one, inflaton field, usually scalar in nature, must be involved.  
We therefore argue the gravitational picture provides a more concrete and natural explanation to the origin and distribution of cosmological matter fluctuations.
Finally, the gravitational fluctuation picture also provides a clear prediction that diverges from scalar field induced predictions on large scales.  
As advanced satellite experiments are continuously being conducted, and increasingly accurate measurements are becoming available, the predictions as a result of quantum gravity made in the last section could be verified or disproved in the near future.

At first it would seem that the assumptions of Gaussian correlations, and therefore of a Gaussian power spectrum, for the scalar inflaton field would be rather restrictive. 
There are in principle many possible scalar field self-interaction terms that can be written 
down in four dimensions, starting with monomials of the fields up to possibly even non-local interactions, all leading potentially to wildly different power spectra. 
Nevertheless Wilson's argument \cite{wil72} (and subsequent related extensive rigorous work) in support of triviality of lambda phi four ($\lambda \phi^4$) theory in and above four spacetime dimensions, based in turn on by now well established field theoretic renormalization group arguments, 
suggests that for a wide class of local scalar self-interacting theories the Gaussian (free field) 
fixed point in the most relevant quartic coupling constant will act as an infrared attractor.  
Here the implication of this deep field-theoretic result is that, for a wide class of scalar self-interacting local theories, the assumption of Gaussianity at large distances is reasonably well justified in terms of rather general renormalization group arguments.
On the basis of these arguments one would then expect that all scalar $n$-point functions should be reasonably well described by their free field expressions, as obtained from the repeated application of Wick's theorem.
At the same time, it should be clear that in the current gravity-based model the long range correlation functions are most certainly {\it not} Gaussian, due to the presence of non-trivial anomalous dimensions.

Specifically, the values for the non-trivial scaling dimensions listed above imply that the 
behavior of gravity and matter $n$-point functions (and their related power spectra) is far 
from their Gaussian, or free field, counterpart.
As an example, note that the two-point function result of Eq.~(\ref{eq:corr_pow1}),
and related to it the universal gravitational scaling dimension $n=1$,
also determines the form of the reduced three-point curvature correlation function \cite{ham17}
\beq
< \sqrt{g} \; R(x_1) \; \sqrt{g} \; R(x_2) \; \sqrt{g} \; R(x_3) >_{c \, R}
\;\; \mathrel{\mathop\sim_{d_{ij} \; \ll \; \xi }} \;\;
{ C_{123}  \over d_{12} \, d_{23} \, d_{31}  } \;\; .
\label{eq:corr_triple}
\eeq
with $C_{123}$ a constant, and relative geodesic distances $d_{ij} = \vert x_i - x_j \vert $, etc.
As before, and again by virtue of the quantum equations of motion, these identities can then 
be related to three-point functions involving the local matter density, or more generally to expressions
involving the trace of the energy momentum tensor.
Proceeding along the same line of arguments, analogous expression can be given for the four-point functions, which again will exhibit a distance dependence unambiguously fixed largely by the scaling dimension of the scalar curvature operator, with non-universal amplitudes.
We note here that the relevance and measurements of nontrivial three- and four-point matter density 
correlation functions in observational cosmology was already discussed in great detail some time ago in \cite{pee93}.
Then the results presented here imply that such higher order $n$-point function could, in the not too distant future, provide additional stringent observational tests on the vacuum condensate picture for quantum gravity, and on the non-trivial gravitational scaling dimensions scenario described here and
in  \cite{ham17}.

We should also say that it is possible for our picture of gravitational fluctuations to even coexists with inflation, with both effects providing contributions to the power spectrum.  
We do not explore this idea here in depth, as the primary aim of this paper is to show that the same power spectrum can be produced independent on inflation, but purely from macroscopic quantum fluctuations of gravity using concrete, well-known, and tested methods for dealing with nonperturbatively renormalizable theories.  
Nevertheless, it is a potential concept for exploration in future work.

The ability to reproduce the cosmological matter power spectrum has long been considered one of the ``major successes'' for inflation-inspired models.  
Although within our preliminary study, further limited by the accuracy of present observational data, it is not yet possible to clearly prove or disprove either idea, the possibility of an alternative explanation without invoking the machinery of inflation suggests that the power spectrum may not be a direct consequence or a solid confirmation of inflation, as some literature may suggest.  
By exploring in more detail the relationship between gravity and cosmological matter and radiation, together with the influx of new and increasing quality observational data, one can hope that this 
hypothesis can be subjected to further stringent physical tests in the near future.

Note added in proof : After this work was completed, the Planck collaboration published an updated power spectrum analysis \cite{akr18} with refined data and error sets, which has led here to the important addition of Figure 6.

\newpage

% \vspace{30pt}

% BELOW BLOCK ONLY USED WHEN USING BIBTEX

%\bibliography{bibv1}        % importing file bibv1.bib
%\bibliographystyle{plain}   % OPTION 1: Choose plain style for bibliography (This one works well the first try.)
%\bibliographystyle{prsty}   % OPTION 2: Choose Phys. Rev. style for bibliography (This one looks weird for some reason on first try.)

% MORE MANUAL BIBLIOGRAPHY USING BIBITEM

\vfill

%\newpage

\end{document}